FAPP and Non-FAPP: A Pedagogical Essay

Jeremy Bernstein

Stevens Institute of Technology

> Might I say immediately…we always have had a great deal of difficulty in understanding the world view that quantum mechanics represents…I cannot define the real problem, therefore I suspect there's no real problem, but I am not sure there's no real problem.
>
> Richard Feynman

1.FAPP

In one of his last essays the late John Bell introduced the very useful notion of FAPP-For All Practical Purposes. I want to distinguish between two kinds of quantum mechanics. There is a kind that I will refer to as FAPP mechanics. This is the kind that you will find occupying most space in most text books on the subject. From these the student will learn about the hydrogen atom, about chemical bonding and maybe even what makes the Sun shine. Any student who wants to become a professional physicist will master FAPP mechanics. It is indispensable and also it is true. Unless such a student is an incipient John Bell he or she will not find any real problems here. The difficulty in understanding usually will have to do with working out the mathematics. So what is Feynman talking about? This is best illustrated by an example which I will give in two parts-FAPP and non-FAPP in that order.

My example involves "spin." Many of you will know what this is but some of you may not.[1] In quantum mechanics all atoms and their constituents have spin. If we want to give a layperson some sense of what this is we describe the motions of the Earth as an example. It goes around the Sun in a nearly circular orbit. Because of this orbital motion is has something called its "orbital

---

[1] I am addressing this essay to non-physicists as well.

angular momentum" which would manifest itself if it ever collided with something. The orbital angular momentum would vanish if the Earth ever came to rest. But even then it might have an additional angular momentum due to the fact that it is revolving around an axis which is what produces night and day. This if you like is the Earth's "spin."  But the quantum mechanical spin differs in that the axis of revolution can only point in a fixed number of directions. It is "quantized." All of the elementary particles have a spin which might be zero. The electron, to take an example, has in physicist's units spin-1/2. This means that its axis can point in only one of two directions-either up or down. This has, as I will now explain, observable consequences. Indeed these consequences were first observed by the German physicists Otto Stern and Walther Gerlach although they did not know that this was what they had observed,

                       The experiments were first done in Germany in the early 1920's. They were not done on electrons but rather on silver atoms. But implicitly, as it turned out, they were done on electrons. This has to do with the atomic structure of silver which Stern and Gerlach did not know. This is depicted below.

Figure1.

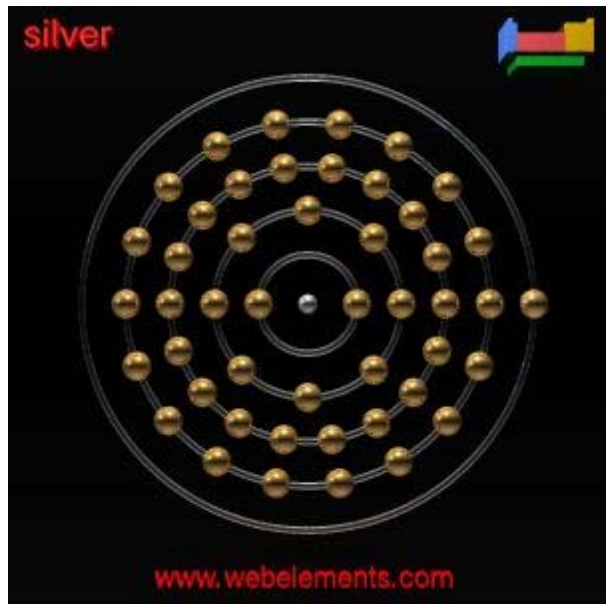

The brass balls represent the electrons and the silver ball in the middle the nucleus. Spatially the diagram is out of whack since the electrons in reality are very far from the nucleus but it does illustrate the point I want to make. Notice that all but one of the electrons are distributed in closed "shells." The outside electron is responsible for chemical bonding since it can be shared with another atom. It is also responsible as it happens for all the angular momentum the silver atom has. What Stern and Gerlach did not know was that this electron has no orbital angular momentum so that the entire angular momentum of the atom is the spin of this electron.

Thanks to the work of Bohr it was generally recognized at this time that angular momentum was "quantized." This meant that the quantity that represented it could only point in a fixed number of directions. Spin was unknown so Bohr was referring to orbital angular momentum. This was the proposition that Stern wanted to test. In principle his idea was very simple. Because of its angular momentum the atom can interact with an external magnetic field. The rotating charged atom acts like a tiny magnet. Suppose one succeeded in producing such a field and suppose it was, say pointing north. For the magnetic effect to work this field would have to vary in strength in space. Then there would be a force acting on the atom and the direction of this force would depend on how the angular momentum was pointing. Thus a beam of such atoms would be separated and the atoms would follow different trajectories depending on how the angular momentum was pointing. If the atoms were then detected on say a photographic plate there should be more than one line on which they appeared if angular momentum was truly quantized. If not there would be simply a continuous smear.

Stern put the matter of constructing the magnetic field to Gerlach and he succeeded. Silver atoms were heated in a furnace and a beam was produced that entered the magnet. Below is a post card that Gerlach sent to Bohr in 1921. The image on the right side shows the split beam. The left image is with the magnetic field turned off so there is no splitting.

Figure 2.

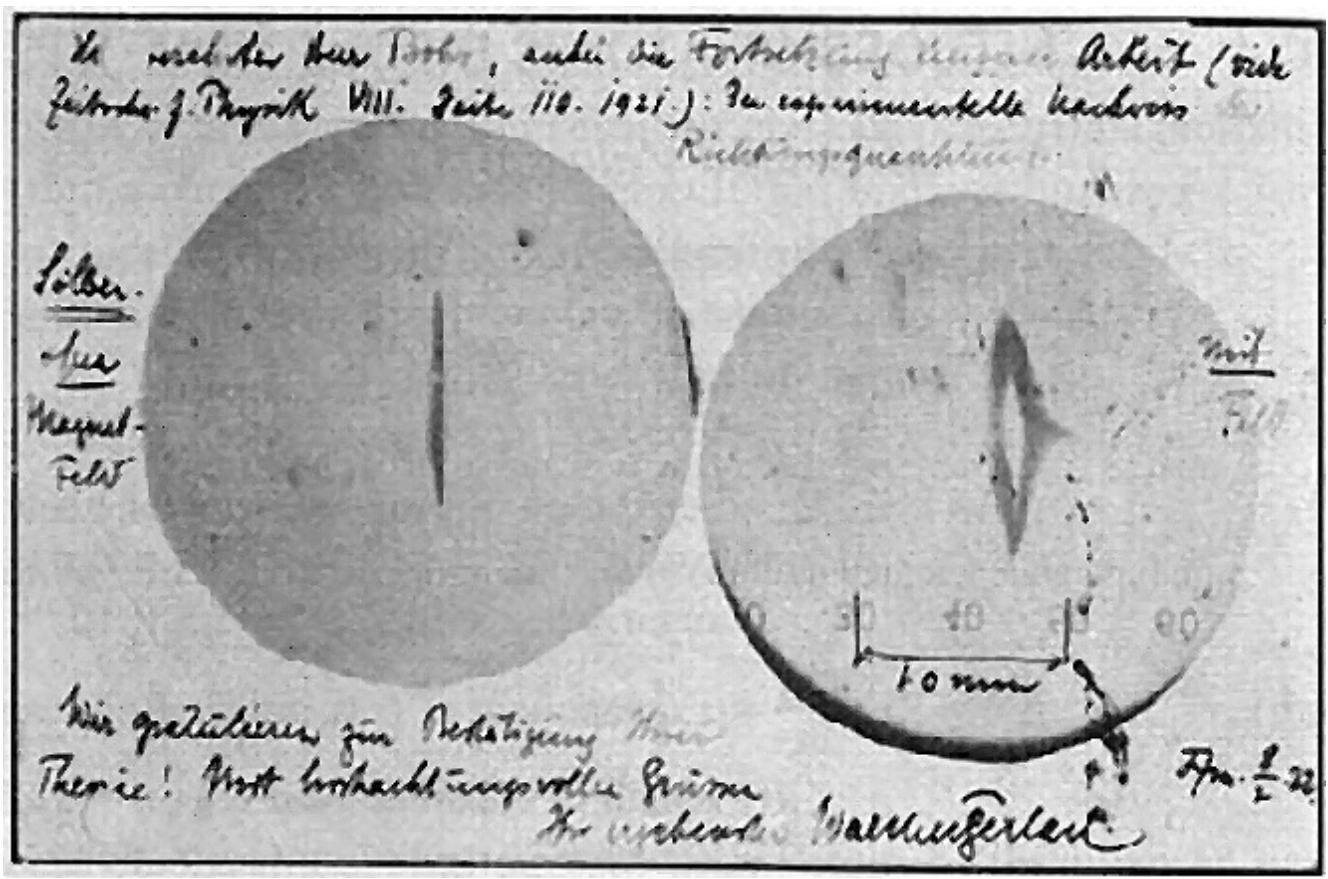

Bohr had some argument that had persuaded him that should be such a splitting but the real explanation in terms of spin was a few years away. However most physicists at the time of the Stern- Gerlach experiment were very impressed. This was a real tangible manifestation of quantization. But it must be emphasized that this-at least what I have said so far- is all in the domain of FAPP mechanics. None of it has anything to do with the sort of thing that Feynman is remarking on. That begins when we consider a setup with two Stern Gerlach magnets or if we consider the measurement process of a single magnet more closely. Below we have a schematic picture of the two magnets with their fields having parallel orientations. Here the two magnets are close to each other but imagine them to be widely separated. What is it that we want to put in the middle?
Figure 3.

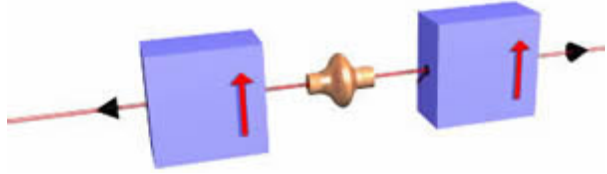

In the original experiment the silver atoms came out of a furnace with their spins randomly oriented. There were as many "up" as down". That is why the two branches in the right picture in the post card above have about the same appearance. But now we want to do something quite different and not readily doable in the form that I am going to describe. The idea of this experiment was first suggested in a 1951 text book by David Bohm.[2] This is one of the few texts that goes beyond FAPP so I put this discussion in the Non-FAPP category.

2 Non-FAPP

"We continue. That a portion of the knowledge should float in the form of disjunctive conditional statements *between* the two systems can certainly not happen if we bring up the two from opposite ends of the world and juxtapose them without interaction. For then indeed the two "know" nothing about each other. A measurement on one cannot possibly furnish any grasp of what is to be expected of the other. Any "entanglement of predictions" that takes place can obviously only go back to the fact that the two bodies at some earlier time formed in a true sense *one* system, that is were interacting, and have left behind *traces* on each other. If two separated bodies, each by itself known maximally, enter a situation in which they influence each other, and separate again, then there occurs regularly that which I have just called *entanglement* of our knowledge of the two bodies. the combined expectation-catalog consists initially of a logical sum of the individual catalogs; during the process it develops causally in accord with known law (there is no question whatever of measurement here). The knowledge remains maximal, but at its end, if the two bodies have again separated, it is not again split into a logical sum of knowledges about the individual bodies. What still remains *of that* may have becomes less than maximal, even very strongly so. --One notes the great difference over against the classical model theory, where of course from known initial states and with known interaction the individual end states would be exactly known." [3] Erwin Schrödinger

---

It is my hope that when you finish reading this section the meaning of this somewhat enigmatic paragraph will become clear or at least clearer to you.

Both FAPP and non-FAPP mechanics have certain iconic concepts in common although they may be examined to different depths. The Heisenberg uncertainty principles are certainly a case in point. In FAPP mechanics we would give examples using such things as the Heisenberg microscope to show by examining the interaction of the light quanta with the object under observation how shortening the wave length of the quanta so as to more accurately determine the position of the object increases the quanta's momentum and thus the object's momentum because of its collision with the light quantum is less and less well-determined. But if we have our non-FAPP hat on we might argue that in the FAPP version it looks as if the object **has** a momentum and somehow we have failed to find a way of measuring it. We would point out that this notion will lead eventually to an inconsistency. The non-FAPPist would argue that without a measurement one cannot speak of the object's momentum. Likewise both the FAPPist and non-FAPPist will consider what Schrödinger calls "entanglement" but will draw different conclusions.

First let me discuss a somewhat fanciful example of classical entanglement. You and I are playing billiards on your table. I have left you with the situation where there are three balls including the cue ball. The two other balls are touching each other in the center of the table. You have the enviable task of knocking one of these balls into a side pocket and hopefully leaving the other in a position where you can pocket it in the next shot thus running the table. Indeed you put one ball in a side pocket. The other skitters off but when the first ball goes into a side pocket the second changes course and deposits itself in the other side pocket. My suspicions are aroused and I ask you to do this again. This time you put the ball in the other pocket and once it is lodged therein the remaining ball again changes course and deposits itself in the pocket opposite. Now I am **really** suspicious and demand an explanation. You say that there is a trick. The table is rigged so that you can change its surface imperceptibly and thus influence the trajectory of the ball. While I may accept this as an explanation it is unlikely that I will play another game of billiards with you on this table. Now to quantum mechanics.

If two spin ½'s are added then the quantum rules tell us that the sum can be either zero or one. To simplify the discussion I will consider the spin zero case. I want to introduce a bit of notation. The object $↑_i$ will represent the ith particle with its spin up, while $↓_i$ will do likewise for spin down. The spin zero state-the so-called spin "singlet"- can be represented by a function that is proportional to $↑_2↓_1-↓_2↑_1$. This is an example of what Schrödinger is calling an "entangled" state. You cannot write it as a simple product of the spin up and down functions for the individual particles. We can contrast this with one of the functions that describe the spin-one situation, say $↑_1↑_2$. Here the spins are not entangled. The entangled state will persist no matter how far the particles are separated provided that neither particle has an interaction with something. This leads, as we shall now see, to what Einstein called "spooky actions at a distance."

In the middle of Figure 3 I imagine a device that manufactures electrons in the singlet state. Moreover these electrons are produced with equal and opposite momenta. There is no definite assignment of a direction of spin to any of these electrons. The spin is entangled. When I measure the spin of the electron that arrives at my Stern-Gerlach magnet sometimes the spin will point up-+- and some times down ---. Thus I will record a random pattern such as +++--+-… . The electrons will also arrive at your Stern-Gerlach magnet and you will also measure an apparently random pattern. We assume that the magnets are so widely separated in space that any communication between them with messages that travel no faster than the speed of light occur only after the fact. But later we can compare notes and we find something quite remarkable. If my pattern is +++--+…then yours is ---++-…. There is a perfect anti-correlation between the spin measurements. It does not matter which one of us measures the spin first, the other one of us will find that in this measurement the spin points on the opposite direction and this persists whatever the spatial separation of the two magnets. If we are a person in the street confronted with this , and not brainwashed by quantum mechanics, the obvious question to ask is how do the spins know how to do this? What is the trick? Where is the table with the rubber surfaces? This is the sort of question that bothered Einstein. If we are Bohr we say there is no "how" and, in fact, you have no right to ask for a "how." This is just they way in which the quantum mechanical cookie crumbles.

I have never been able to find in anything that I have read of Einstein I which he states exactly what sort of explanation would have satisfied him. There is one thing that we know would not have satisfied him which we can call loosely a quantum theory with hidden variables. I will now try to explain what such a theory is. In 1924 Louis de Broglie introduced in his *these* –a more advanced French version of a PhD thesis-the notion that electrons, say, could manifest wave-like characteristics as well as particle characteristics. A few years after quantum mechanics was developed de Broglie presented an idea for how this particle-wave duality might be understood. In his proposal there would be particles and these would follow classical trajectories guided by "pilot" waves which satisfied Schrödinger's wave equation. When he presented this notion it was roundly criticized by people like Pauli and de Broglie dropped it. It was picked up in the early 1950's by the aforementioned David Bohm. There is a curious irony in this since in 1951, as I have mentioned, he published his quantum mechanics book in which he made a strong argument against the possibility of any classical theory of this type reproducing all the results of the quantum theory. He wrote, "We conclude then that no theory of mechanically determined hidden variables would agree with quantum theory for a wide range of predicted experimental results." [4] Then he appends a curious footnote," We do not wish to imply that anyone has ever produced a concrete and successful example of such a theory, but only state that such a theory is, as far as we know, conceivable."[5] What is curious is that at about the same time that Bohm was writing this he had produced just such an example.

---

[4] Bohm, op cit p,623.

We come to it next but I want first to discuss von Neumann. In 1932 he published his masterpiece *Matematische Grundlagen der Quantenmechanik*. An English translation under the title "Mathematical Foundations of Quantum Mechanics" was published in 1955.[6] In the book, as the title suggests, von Neumann analyses the mathematical structure of the theory. He also analyses how measurements are performed and what interests us can the statistical results of the theory be explained by what he calls "hidden parameters" which are classical. An analogy, at least in sprit, is the relationship between thermodynamics and statistical mechanics. Here the hidden parameters are the momenta say of the molecules of a gas. We can study the thermodynamic laws of the gas without any knowledge of these. They are hidden parameters. Von Neumann argues that for quantum mechanics this is impossible. I at least found his argument as presented very difficult to understand. It only became clear to me when I read an explanation by the aforementioned John Bell.[7]

Von Neumann notes that if there was such a possibility and if these parameters were classical there would have to be what he calls "dispersion free states." What characterizes classical physics is that the parameters such as position and momentum can be jointly specified with arbitrary precision. But in quantum physics they cannot be and the best one can do is to present the probable values that such observables might have always respecting the Heisenberg uncertainty relations. But given such a spectrum of possibilities one can always ask what is the value one might expect if one did many measurements. If such an observable is called A then this expected value can be

---

[5] Bohm, op. cit. p.623.
[6] Princeton University Press, Princeton, New Jersey, 1955.
[7] John Bell, Speakable and Unspeakable in Quantum Mechanics, Cambridge University Press, Cambridge, 2004, See the discussion on page 4,

called $\langle A \rangle$. If we have two such observable A and B, even if they are subject to an uncertainty relation between them, then the quantum theory tells us that

$\langle A+B \rangle = \langle A \rangle + \langle B \rangle$. If one thinks about it, this is certainly a curious result since A and B may be like position and momentum, observables with a tension between them,[8] but this is what the theory tells us. But the hypothetical dispersion free state is special. In this state the expected values are simply one of the possible results of a measurement with no dispersion. Thus the sum of the expected values would simply in this case be the sum of these measurement results. But this equality is not generally true .A possible value of (A+B) is not in general the sum of possible values of A and B.[9] Bell gives a simple example involving spins which demonstrates this. Hence von Neumann argues there can be no dispersion free states in quantum mechanics. His argument is correct but as you will now see it has nothing to do with Bohm's version of quantum mechanics.

Indeed Bohm's version is almost the inverse of von Neumann's. In the case of Bohm it is the quantum mechanical object that is hidden while the classical one is out in the open. In Bohm, as in FAPP mechanics, one begins with the Schrödinger equation. The solutions to this equation are the "wave functions" which in FAPP mechanics are used to deduce the probable outcomes of possible experiments. In Bohm's model these same solutions serve an entirely different purpose. They are pilot waves of the classical particles. There is an equation for the position of such a particle as it evolves on time and this evolution is determined by a pilot wave which is in turn determined by the Schrödinger equation. Let is consider, at least qualitatively an example.

One of the most iconic situations in FAPP mechanics is what is called the "two-slit" experiment. A barrier is erected but it has two narrow slits that can be open or closed. Behind the barrier there is particle detector-say a

---

[8] Mathematically they may not commute.
[9] I have tried to avoid the use of the term "eigen-value" so as not to burden a reader unfamiliar with this expression. But in that language what I am saying is that eigen values of the sum of two non-commuting operators is not in general the sum of the eigen values.

photographic plate. To make the situation as paradoxical as possible I will suppose that a beam of, say electrons, is prepared of such a character that the electrons are allowed to approach the barrier one at a time. Actually experiments like this have been done. Below are the results of such an experiment done by the Japanese physicist Akira Tonomura and his collaborators,

Figure 4.

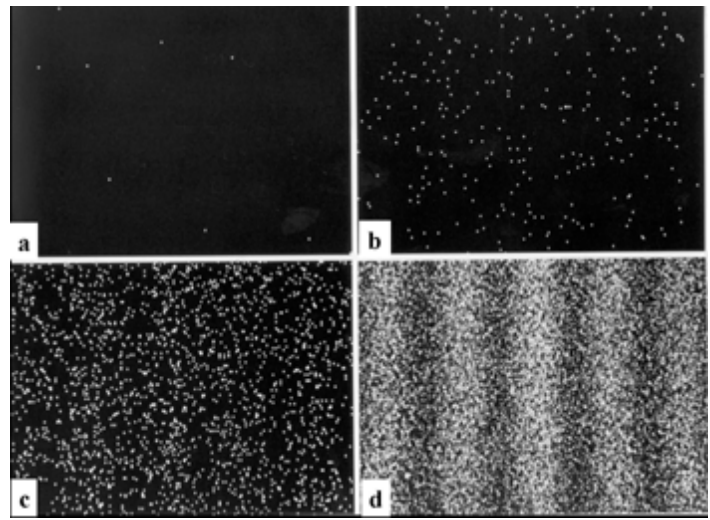

What do these pictures mean?  In this experiment both slits are open. If we were talking about classical waves the waves going through the slits would interfere with each other and produce interference effects such as are clear in the picture d above. But this is not what is happening here. The electrons, which have a wave as well as a particle nature, are going through **one at a time** but over the course of time build up the same pattern as if they were somehow going through both slits at once. To make matters worse, if one of the slits is closed the interference fringes in 'd' disappear and are replaced by a single blob. Again someone who has not been brainwashed by FAPP mechanics would ask how does the electron "know" that the other slit  is open or  closed ? He or she would be told by Bohr that this is a silly question. The formalism of FAPP mechanics predicts this behavior and that is all you have to know. How is the same phenomenon viewed in Bohm's mechanics? In principle this is quite simple. The pilot wave goes through both slits and thus knows if a slit is open or closed but the electron only goes through one guided by the pilot wave. Below is a result of a calculation of these waves. Where the density of the lines is greatest is where the electron will be most likely guided to do.

Figure 5.

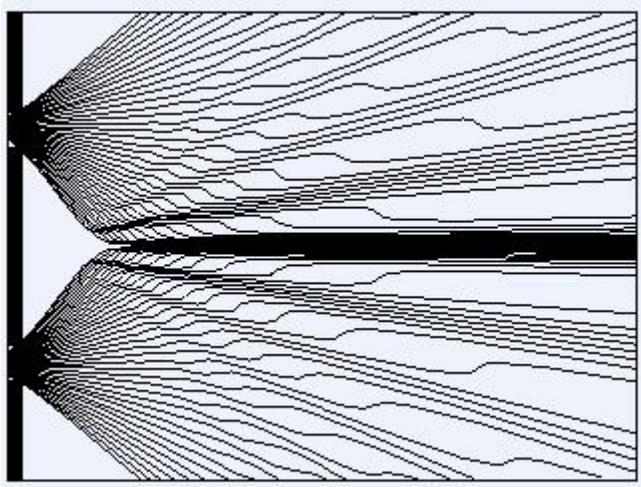

Many physicists-probably most-would argue that this game is not worth the candle. One is much better off to accept the FAPP result and be done with it. And yet…

In any event the complexity of the Bohm model is not here. One must keep in mind that this model is not relativistic yet it should not contradict relativity. This enters when we consider two particles in interaction. Once again we begin by solving the Schrödinger equation. Now we run into the matter of entanglement. The solutions, just as in our spin case, cannot be factored into a product belonging to one particle times a product belonging to the other. But the situation is more complex. The solution depends on data involving both particles collectively. In other words the theory is "non-local." An argument can be made that in this case the non-locality does not involve any exchange of signals moving at speeds greater than the speed of light. But this non-locality is what allows the theory to describe the correlations we discussed earlier, The pilot waves acquire data from both Stern-Gerlach magnets and guide the electrons accordingly so as to reproduce the anti-correlation which quantum mechanics predicts. When Bell realized this he asked whether one could find another model with some of these features which was both local and reproduced all the results of the quantum theory. He found that he could not and the experiments confirm this. The non-locality which Einstein called the "spooky actions at a distance" simply must be accepted in any domain in which the quantum theory is valid. But this raises the question, in what domains are these?

For decades Einstein and Bohr battled over the soul of the quantum theory. Bohr won all the battles handily but he may have lost the war. Bohr insisted that experiments must be described classically. This meant that there had to be a division between the quantum and classical worlds. But he was never able to specify exactly what this division was. Of course FAPP such divisions can be made, but this is hardly satisfactory. More modern approaches argue that there is no division. The whole world is quantum mechanical although FAPP we often need not take this into account. So far I have not seen an example of such an approach that I am entirely comfortable with. They generally have a probabilistic view of the past. Many pasts are possible but some are more probable than others. To me the past is classical. There is only one. I suspect that Einstein was right. Quantum mechanics is only part of the story. To those who think that it is the whole story I am reminded of something Schrödinger said the one time I visited him with colleagues in his apartment in Vienna. The apartment building incidentally looked like something out of The Third Man- elevator and all. His apartment was stacked with books in every possible language. There was no cat. He did not like cats. He wanted to know what physicists were working on. We told him and as we were leaving he said, " There is one thing we have forgotten since the Greeks-modesty!"